\documentclass[runningheads]{llncs}
\usepackage[T1]{fontenc}
\usepackage{tabularx}

\usepackage{graphicx}
\usepackage{orcidlink}
\begin{document}
\title{Cold Start Problem: An Experimental Study of Knowledge Tracing Models with New Students}
\titlerunning{Cold Start Problem in Knowledge Tracing}
%
\author{Indronil Bhattacharjee \orcidlink{0000-0002-3463-3389} \and
Christabel Wayllace \orcidlink{0000-0001-8039-2777}}
\authorrunning{I. Bhattacharjee and C. Wayllace}
%
\institute{New Mexico State University, Las Cruces, New Mexico, USA \\
\email{\{indronil, cwayllac\}@nmsu.edu}}

\maketitle
%
\begin{abstract}
 Knowledge Tracing (KT) involves predicting students' knowledge states based on their interactions with Intelligent Tutoring Systems (ITS). A key challenge is the cold start problem—accurately predicting knowledge for new students with minimal interaction data. Unlike prior work, which typically trains KT models on initial interactions of all students and tests on their subsequent interactions, our approach trains models solely using historical data from past students, evaluating their performance exclusively on entirely new students. We investigate cold start effects across three KT models: Deep Knowledge Tracing (DKT), Dynamic Key-Value Memory Networks (DKVMN), and Self-Attentive Knowledge Tracing (SAKT), using ASSISTments 2009, 2015, and 2017 datasets. Results indicate all models initially struggle under cold start conditions but progressively improve with more interactions; SAKT shows higher initial accuracy yet still faces limitations. These findings highlight the need for KT models that effectively generalize to new learners, emphasizing the importance of developing models robust in few-shot and zero-shot learning scenarios.

\keywords{Knowledge Tracing  \and Cold Start Problem \and Deep Learning \and Educational Data Mining.}
\end{abstract}
\section{Introduction}

Intelligent Tutoring Systems (ITS) aim to provide personalized learning by adapting to students' evolving knowledge. Knowledge tracking (KT), which determines student knowledge from interactions, is crucial but faces the cold start problem, where models struggle with minimal initial data, delaying effective personalization.

Cold start in KT has been studied across various scenarios, including cold start questions \cite{guo:24}, cold start skills \cite{coldstart_skill}, and new activities \cite{zhao:20}. Typically, KT models are trained on initial student interactions and evaluated on subsequent interactions \cite{dkt,khajah,dkvmn,sakt,akt,skvmn,rkt,ekt}, thus not fully assessing their performance on entirely new students.

This study examines the cold start issue in three KT models—DKT \cite{dkt}, DKVMN \cite{dkvmn}, and SAKT \cite{sakt}—by training exclusively on historical student data and testing solely on unseen students using three datasets (ASSISTments 2009 \cite{assist2009}, 2015 \cite{assist2015}, and 2017 \cite{assist2017}). Our contributions include: (1) a realistic evaluation framework, (2) a comparative analysis of KT model adaptability, and (3) empirical evidence showing that, despite better initial accuracy, attention-based architectures still face cold start limitations, highlighting a need for improved model generalization.

\section{Related Research}

Intelligent tutoring systems \cite{its} and personalized learning environments have significantly advanced through extensive research on student modeling and knowledge tracing. Early work by Piech et al. (2015)~\cite{dkt} introduced Deep Knowledge Tracing (DKT), utilizing recurrent neural networks (RNNs) \cite{rnn} to capture student learning dynamics. Addressing DKT’s limitations in interpretability and long-term dependency handling, subsequent models like Dynamic Key-Value Memory Networks (DKVMN) \cite{dkvmn} and Self-Attentive Knowledge Tracing (SAKT) \cite{sakt} emerged, using external memory and attention mechanisms, respectively, for enhanced performance \cite{khajah,akt,skvmn,rkt,ekt}.

A critical issue in knowledge tracing is the cold start problem, accurately predicting knowledge states for new students with limited data, highlighted extensively in educational \cite{its} and recommender systems literature \cite{rec,rec_sys}. Studies addressing cold start scenarios often use hybrid methods \cite{hybrid}, auxiliary data \cite{data_length}, or sparse data robustness \cite{sparse}. Zhang et al. (2017) \cite{coldstart_skill} analyzed skill-based cold start conditions, showing initial predictive advantages that diminish over subsequent interactions. Recent approaches specifically targeting the cold start problem include Zhao et al.’s (2020) differential memory-based model \cite{zhao:20} and Jung et al.’s (2024) generative language model framework (CLST) \cite{jung}. Unlike these skill-oriented studies, our research specifically evaluates generalization to entirely new students without prior interaction data, emphasizing a broader adaptation challenge.

\section{Problem Statement and Research Approach}
As described earlier, a key challenge in KT is the cold start problem, where models must predict a new student’s knowledge state with little to no prior data. Traditional KT models rely on sequential patterns, but minimal interaction data reduces accuracy, leading to unreliable early recommendations and potentially suboptimal learning experiences.

\subsection{Our Research Approach}
Unlike prior work that evaluates KT models using a student’s own initial responses \cite{dkt,dkvmn,sakt,skvmn,akt,ekt}, our study investigates the cold start effect by training models on historical data from past students and testing them on entirely new learners. This approach better simulates real-world ITS deployment, where models must generalize beyond previously seen individuals. We conduct experiments on three widely used KT models—DKT \cite{dkt}, DKVMN \cite{dkvmn}, and SAKT \cite{sakt}—across multiple datasets (ASSISTments 2009, 2015, and 2017) to systematically evaluate their performance in cold start scenarios.
\subsection{Key Research Questions}
Our study aims to address the following research questions:
\begin{enumerate}
    \item How do KT models perform when predicting knowledge states for new students with minimal prior data?
    \item Which KT models adapt more effectively as students engage with additional learning exercises?
    \item Does an attention-based approach, such as SAKT, inherently mitigate the cold start problem?
\end{enumerate}

\section{Experiment setup and methodology}
To evaluate cold start performance, we set our experiment by training KT models on historical data from previous students and tested them on five randomly selected sets of new students, ensuring complete data segregation. We then measured model performance over increasing interactions to capture their learning curves and adaptability in real-world ITS scenarios.

\subsection{Datasets}
This study utilizes three widely recognized datasets from the ASSISTments platform \cite{assistment}, known for providing comprehensive benchmarks in Knowledge Tracing (KT). These datasets comprise student interactions with mathematical exercises from the Massachusetts Comprehensive Assessment System (MCAS) \cite{ghodai}. The ASSISTments2009 dataset \cite{assist2009}, from the 2009-2010 academic year, is employed in its refined 'skill-builder' version, which resolves earlier data issues and includes detailed knowledge component (KC) annotations. The Assist2015 dataset \cite{assist2015} from 2015 provides fewer attributes and lacks KC annotations, creating challenges due to its limited context. Lastly, Assist2017 \cite{assist2017}, also known as the ASSISTment Challenge dataset, spans data from 2004 to 2006 and offers extensive metadata and 102 KCs, suitable for deeper analyses of complex student interactions.

\begin{table}[ht]
\centering
\caption{\textbf{Overview of ASSISTments Datasets}}
\label{table:datasets}
\begin{tabular}{|p{0.75in}|p{0.75in}|p{0.75in}|p{0.75in}|p {0.75in}|p {0.75in}|}
\hline
\textbf{Dataset} & \textbf{Year(s)} & \textbf{Interactions} & \textbf{Students} & \textbf{Questions} & \textbf{KCs}  \\ \hline
Assist2009 & 2009-2010 & 346,860 & 4,217 & 26,688 & 123 \\ \hline
Assist2015 & 2015 & 708,631 & 19,917 & 100 & - \\ \hline
Assist2017 & 2004-2006 & 942,816 & 1,709 & 3,162 & 102 \\ \hline
\end{tabular}
\end{table}

\subsection{Test Set Preparation and Data Segregation}
To rigorously evaluate KT models under realistic cold start conditions, we randomly selected five distinct sets of 10 students from each dataset, ensuring diverse academic behaviors and proficiency levels. We then completely removed these students' prior interaction data from the training sets, simulating scenarios where models encounter entirely new learners without prior knowledge of their performance. The remaining training data preserved comprehensive interaction diversity, ensuring sufficient variability for effective generalization during model training. This approach allowed us to precisely assess each model's capability to adapt and accurately predict knowledge states for new students.

\subsection{Models}
\textbf{Deep Knowledge Tracing (DKT)} \cite{dkt} uses recurrent neural networks (RNNs) \cite{rnn}, particularly LSTM \cite{lstm}, to capture temporal dependencies in student interactions, enabling sequential predictions of student performance. \textbf{Dynamic Key-Value Memory Networks (DKVMN)} \cite{dkvmn} employs an external memory structure with separate keys (concept representations) and values (student mastery levels), dynamically updating memory based on incoming interactions. \textbf{Self-Attentive Knowledge Tracing (SAKT)} \cite{sakt} uses attention mechanisms inspired by Transformer architectures, directly modeling relationships among past interactions and focusing on the most relevant events for predicting student knowledge states.

\subsection{Experiment Configuration}
The experimental setup assesses KT model adaptation to new students across increasing interaction sequences. For DKT and DKVMN, experiments cover sequences from 3 to 20 questions, examining initial adaptability and performance progression. For SAKT, due to its attention-based architecture designed to handle longer interaction sequences effectively, we extended the evaluation up to 30 questions to determine if it maintains superior performance over prolonged interactions.

\section{Results}

We evaluated how DKT, DKVMN, and SAKT adapt to new students under cold start conditions across the ASSISTments 2009, 2015, and 2017 datasets. Each model was tested on five student sets with interaction sequences increasing from 3 to 20 (30 for Assist2017), simulating real-world deployment for unseen learners.


\textbf{DKT} shows steady improvement across all datasets as more student interactions become available. In Assist2009 and 2015, accuracy consistently rises from ~0.45–0.60 to ~0.75 by the 20th question, especially for Sets 1 and 2. Despite some fluctuation in Sets 4 and 5, the model demonstrates strong late-stage adaptation. Assist2017 highlights DKT's ability to learn over longer sequences, eventually matching or exceeding early-leader sets despite slower initial gains. (Fig. 1a). \textbf{DKVMN} consistently performs well across all datasets. It achieves high initial accuracy, particularly in Sets 1 and 2, and rapidly improves to over 0.75 in most conditions. In Assist2015, DKVMN outperforms the other models, showing sharp gains even in Set5. Its memory-based architecture enables effective prediction even under limited data, with stable performance across longer sequences in Assist2017. (Fig. 1b)

\begin{figure}[ht] \centering 
\includegraphics[width=0.41\linewidth]{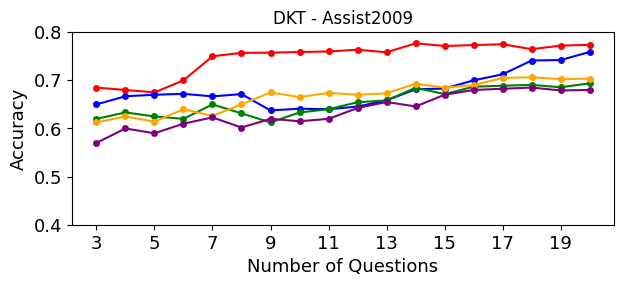} \includegraphics[width=0.41\linewidth]{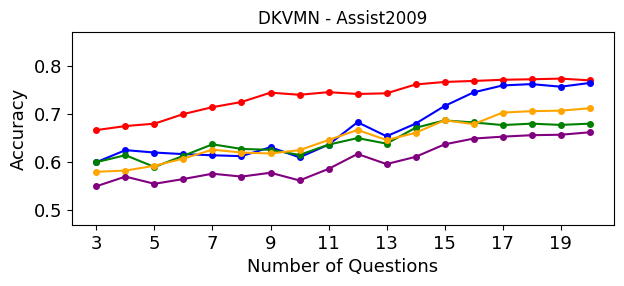} \includegraphics[width=0.41\linewidth]{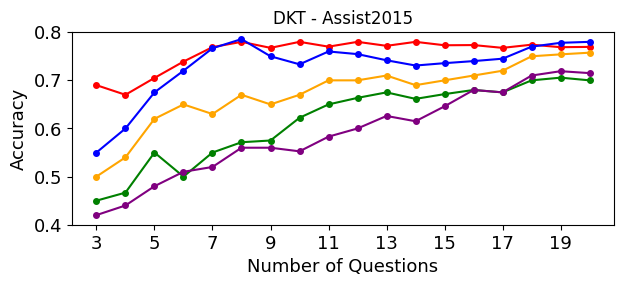}
\includegraphics[width=0.41\linewidth]{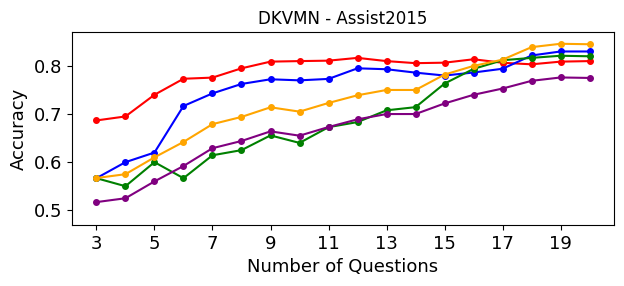} \includegraphics[width=0.41\linewidth]{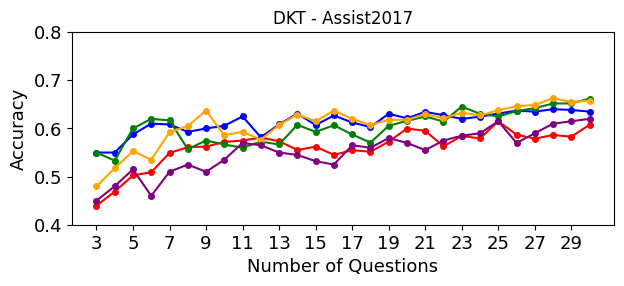} \includegraphics[width=0.41\linewidth]{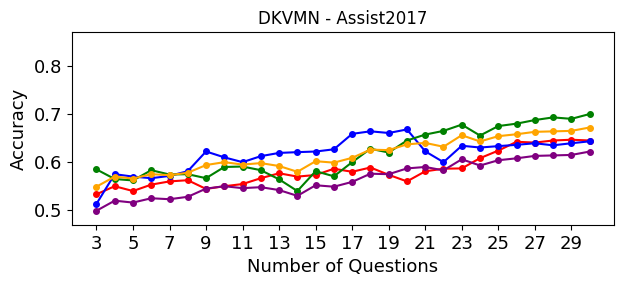}
\includegraphics[width=0.65\linewidth]{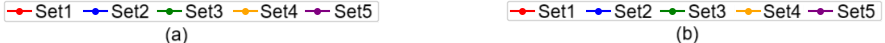}
\caption{Accuracy vs Number of Questions (2009, 2015, 2017) - (a)DKT (b)DKVMN} 
\end{figure}

\begin{figure}[ht] \centering 
\includegraphics[width=0.41\linewidth]{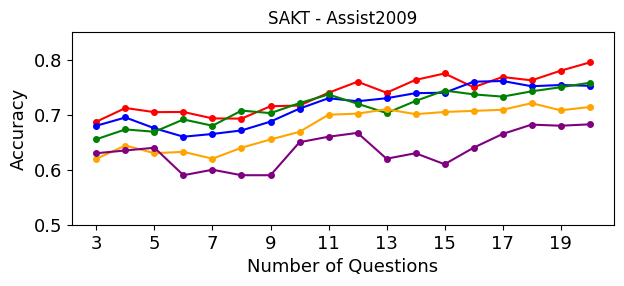} 
\includegraphics[width=0.41\linewidth]{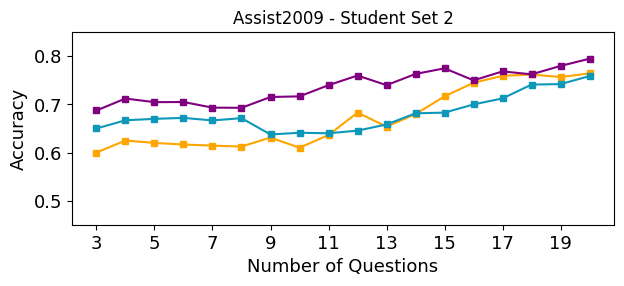} 
\includegraphics[width=0.41\linewidth]{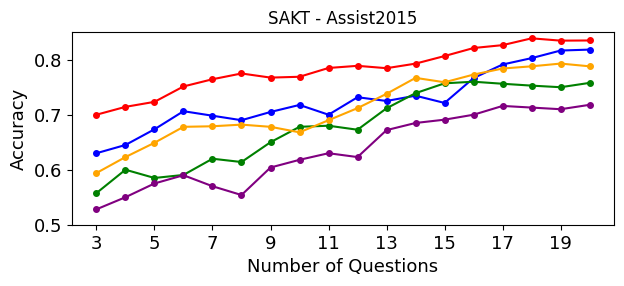} 
\includegraphics[width=0.41\linewidth]{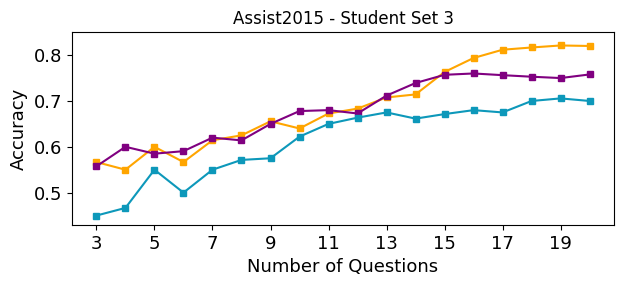} 
\includegraphics[width=0.41\linewidth]{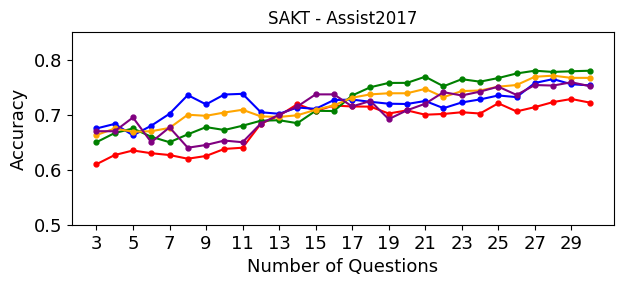} 
\includegraphics[width=0.41\linewidth]{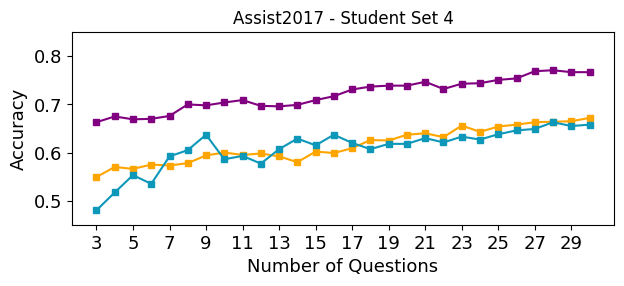}
\includegraphics[width=0.65\linewidth]{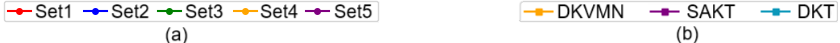}\caption{(a) SAKT Accuracy vs Number of Questions (2009, 2015, 2017)   and  
(b) Accuracy vs Number of Questions with Different Models and Different Student Sets} \end{figure}

\textbf{SAKT} shows quick initial adaptation in Assist2009 and 2015, often outperforming other models early on. Accuracy rises steadily across all sets, with Sets 1–3 reaching competitive levels by the end. In Assist2017, performance improves gradually across the extended sequence length, though some sets show mild plateaus in later stages indicating a need for enhanced long-range attention. (Fig. 2a). The comparison across different student sets and datasets highlights the varying strengths and weaknesses of each KT model in adapting to new student interactions. (Fig. 2b).

\section{Discussion}
The results provide insight into the research questions. For RQ1, all KT models show initial difficulty in accurately predicting knowledge states for new students, confirming the impact of limited prior data on early model performance. Regarding RQ2, DKVMN adapts fastest with minimal data, while DKT improves steadily as more interactions become available, indicating its strength in long-term adaptation. For RQ3, although SAKT shows better early performance due to its attention mechanism, it often plateaus, suggesting that attention alone does not inherently solve the cold start problem.

\section{Conclusion and Future Work}
The cold start problem poses a critical challenge for KT models, especially when predicting the knowledge states of entirely new students. By training models solely on historical student data, our study simulated realistic cold start conditions and revealed distinct behaviors across three KT models. DKVMN excelled in early predictions due to its memory mechanisms, DKT showed strong sequential adaptability over time, and SAKT leveraged attention for quick gains but often plateaued. These findings underscore the need for models that combine rapid adaptation with sustained learning.

Our future research will focus on developing hybrid KT models that can enhance the unique advantages of existing approaches while mitigating their limitations. Additionally, exploring advanced machine learning techniques such as transfer learning and federated learning could further enhance the adaptability of KT models to new student data. These advancements could lead to more personalized, effective and scalable adaptive learning systems capable of supporting diverse and evolving educational needs.
\bibliographystyle{splncs04}

\begin{thebibliography}{99}

\bibitem{akt}
Aritra Ghosh, Neil Heffernan, and Andrew S. Lan.
\newblock \emph{Context-aware attentive knowledge tracing}.
In Proceedings of the 26th ACM SIGKDD International Conference on Knowledge Discovery \& Data Mining, pages 2330--2339, 2020.

\bibitem{assistment}
The ASSISTments Foundation.
\newblock \emph{ASSISTments | Free Education Tool for Teachers \& Students --- new.assistments.org}.
\newblock \url{https://new.assistments.org/}, [Accessed 20-02-2025].

\bibitem{assist2009}
Shun Mao.
\newblock \emph{Assistment2009}.
IEEE Dataport, 2024. DOI: \url{10.21227/k80b-0n66}.

\bibitem{assist2015}
Douglas Selent, Thanaporn Patikorn, and Neil Heffernan.  
\newblock \emph{Assistments dataset from multiple randomized controlled experiments}.  
In Proceedings of the Third (2016) ACM Conference on Learning@ Scale, pages 181--184, 2016.

\bibitem{assist2017}
Thanaporn Patikorn, Neil T. Heffernan, and Ryan S. Baker.  
\newblock \emph{Assistments longitudinal data mining competition 2017: A preface}.  
In Proceedings of the Workshop on Scientific Findings from the ASSISTments Longitudinal Data Competition, International Conference on Educational Data Mining, 2018.

\bibitem{coldstart_skill}
Jiayi Zhang, Rohini Das, Ryan Baker, and Richard Scruggs.  
\newblock \emph{Knowledge tracing models’ predictive performance when a student starts a skill}.  
In Proceedings of the 14th International Conference on Educational Data Mining. EDM, Paris, France, pages 625--629, 2021.

\bibitem{das:21}
Rohini Das, Jiayi Zhang, Ryan S. Baker, and Richard Scruggs.  
\newblock \emph{A New Interpretation of Knowledge Tracing Models' Predictive Performance in Terms of the Cold Start Problem.}  
In EDM (Workshops), 2021.

\bibitem{data_length}
Moyu Zhang, Xinning Zhu, Chunhong Zhang, Feng Pan, Wenchen Qian, and Hui Zhao.  
\newblock \emph{No Length Left Behind: Enhancing Knowledge Tracing for Modeling Sequences of Excessive or Insufficient Lengths}.  
In Proceedings of the 32nd ACM International Conference on Information and Knowledge Management, pages 3226--3235, 2023.

\bibitem{dkt}
Chris Piech, Jonathan Bassen, Jonathan Huang, Surya Ganguli, Mehran Sahami, Leonidas J. Guibas, and Jascha Sohl-Dickstein.  
\newblock \emph{Deep knowledge tracing}.  
Advances in neural information processing systems, vol. 28, 2015.

\bibitem{dkvmn}
Jiani Zhang, Xingjian Shi, Irwin King, and Dit-Yan Yeung.  
\newblock \emph{Dynamic key-value memory networks for knowledge tracing}.  
In Proceedings of the 26th international conference on World Wide Web, pages 765--774, 2017.

\bibitem{ekt}
Qi Liu, Zhenya Huang, Yu Yin, Enhong Chen, Hui Xiong, Yu Su, and Guoping Hu.  
\newblock \emph{Ekt: Exercise-aware knowledge tracing for student performance prediction}.  
IEEE Transactions on Knowledge and Data Engineering, vol. 33, no. 1, pages 100--115, 2019. IEEE.

\bibitem{hybrid}
Andrea Zanellati, Daniele Di Mitri, Maurizio Gabbrielli, and Olivia Levrini.  
\newblock \emph{Hybrid models for knowledge tracing: A systematic literature review}.  
IEEE Transactions on Learning Technologies, 2024. IEEE.

\bibitem{its}
John R. Anderson, C. Franklin Boyle, and Brian J. Reiser.  
\newblock \emph{Intelligent tutoring systems}.  
Science, vol. 228, no. 4698, pages 456--462, 1985. American Association for the Advancement of Science.

\bibitem{ghodai}
Ghodai Abdelrahman, Qing Wang, and Bernardo Nunes.  
\newblock \emph{Knowledge tracing: A survey}.  
ACM Computing Surveys, vol. 55, no. 11, pages 1--37, 2023. ACM New York, NY.

\bibitem{guo:24}
Yuxiang Guo, Shuanghong Shen, Qi Liu, Zhenya Huang, Linbo Zhu, Yu Su, and Enhong Chen.  
\newblock \emph{Mitigating Cold-Start Problems in Knowledge Tracing with Large Language Models: An Attribute-aware Approach}.  
In Proceedings of the 33rd ACM International Conference on Information and Knowledge Management, pages 727--736, 2024.

\bibitem{khajah}
Mohammad Khajah, Robert V. Lindsey, and Michael C. Mozer.  
\newblock \emph{How Deep is Knowledge Tracing?.}  
International Educational Data Mining Society, 2016. ERIC.

\bibitem{lstm}
S. Hochreiter.  
\newblock \emph{Long Short-term Memory}.  
Neural Computation MIT-Press, 1997.

\bibitem{rec}
Paul Resnick and Hal R. Varian.  
\newblock \emph{Recommender systems}.  
Communications of the ACM, vol. 40, no. 3, pages 56--58, 1997. ACM New York, NY, USA.

\bibitem{rec_sys}
Prem Melville and Vikas Sindhwani.  
\newblock \emph{Recommender systems.}  
Encyclopedia of machine learning, vol. 1, pages 829--838, 2010.

\bibitem{rkt}
Shalini Pandey and Jaideep Srivastava.  
\newblock \emph{RKT: relation-aware self-attention for knowledge tracing}.  
In Proceedings of the 29th ACM international conference on information \& knowledge management, pages 1205--1214, 2020.

\bibitem{rnn}
David E. Rumelhart, Geoffrey E. Hinton, and Ronald J. Williams.  
\newblock \emph{Learning internal representations by error propagation, parallel distributed processing, explorations in the microstructure of cognition, ed. de rumelhart and j. mcclelland. vol. 1. 1986}.  
Biometrika, vol. 71, no. 599-607, pages 6, 1986.

\bibitem{sakt}
Shalini Pandey and George Karypis.  
\newblock \emph{A self-attentive model for knowledge tracing}.  
arXiv preprint arXiv:1907.06837, 2019.

\bibitem{skvmn}
Ghodai Abdelrahman and Qing Wang.  
\newblock \emph{Knowledge tracing with sequential key-value memory networks}.  
In Proceedings of the 42nd international ACM SIGIR conference on research and development in information retrieval, pages 175--184, 2019.

\bibitem{sparse}
Shuyan Huang, Zitao Liu, Xiangyu Zhao, Weiqi Luo, and Jian Weng.  
\newblock \emph{Towards robust knowledge tracing models via k-sparse attention}.  
In Proceedings of the 46th International ACM SIGIR Conference on Research and Development in Information Retrieval, pages 2441--2445, 2023.

\bibitem{zhao:20}
Jinjin Zhao, Shreyansh Bhatt, Candace Thille, Neelesh Gattani, and Dawn Zimmaro.  
\newblock \emph{Cold start knowledge tracing with attentive neural turing machine}.  
In Proceedings of the Seventh ACM Conference on Learning@ Scale, pages 333--336, 2020.

\bibitem{jung}
Heeseok Jung, Jaesang Yoo, Yohaan Yoon, and Yeonju Jang.  
\newblock \emph{CLST: Cold-Start Mitigation in Knowledge Tracing by Aligning a Generative Language Model as a Students' Knowledge Tracer}.  
arXiv, 2024. \url{https://arxiv.org/abs/2406.10296}. arXiv.org perpetual, non-exclusive license.

\end{thebibliography}

\end{document}